\documentclass[fleqn,usenatbib]{mnras}
\usepackage{lscape}

\usepackage[T1]{fontenc}

\DeclareRobustCommand{\VAN}[3]{#2}
\let\VANthebibliography\thebibliography
\def\thebibliography{\DeclareRobustCommand{\VAN}[3]{##3}\VANthebibliography}

\usepackage{adjustbox}
\usepackage{xcolor}

\usepackage{float}
\usepackage{graphicx}	% Including figure files
\usepackage{amsmath}	% Advanced maths commands
\usepackage{amssymb}	% Extra maths symbols
\usepackage{hyperref}
\usepackage{newtxtext,newtxmath}

\newcommand{\Gizmo}{\texttt{Gizmo}}
\newcommand{\Arepo}{\texttt{Arepo}}
\newcommand{\SpherIC}{\texttt{SpherIC}}
\newcommand{\sig}{\,$\text{cm}^{2}\, \text{g}^{-1}$}	% cm-squared per gram
\definecolor{black}{HTML}{000000}
\definecolor{darkblue}{HTML}{226E9C}
\definecolor{lightblue}{HTML}{9EC9E2}
\definecolor{lightgreen}{HTML}{9CCB86}

\newcommand\crule[3][black]{\textcolor{#1}{\rule{#2}{#3}}}

\title[SIDM Code Comparison]{Comparing Implementations of Self-Interacting Dark Matter in the Gizmo and Arepo Codes}

\author[Meskhidze et al.]
{\parbox{17.5cm}{
Helen Meskhidze,$^{1}$\thanks{E-mail: emeskhid@uci.edu} Francisco J. Mercado,$^{2}$ Omid Sameie,$^{3}$ Victor H. Robles,$^{4}$ James S. Bullock,$^{2}$ Manoj Kaplinghat,$^{2}$ James O. Weatherall$^{1}$}\vspace{0.3cm}\\
$^{1}$Department of Logic and Philosophy of Science, 779 Social Science Tower, University of California Irvine, CA 92697, USA\\
$^{2}$Center for Cosmology, Department of Physics and Astronomy, 4129 Reines Hall, University of California Irvine, CA 92697, USA\\
$^{3}$Department of Astronomy, The University of Texas Austin, 2515 Speedway, Stop C1400, Austin, TX 78712 USA\\
${^4}$Physics Department, Yale Center for Astronomy and Astrophysics, New Haven, CT 06520, USA\\
}

\date{Accepted XXX. Received YYY; in original form ZZZ}
\pubyear{2022}

\begin{document}
\label{firstpage}
\pagerange{\pageref{firstpage}--\pageref{lastpage}}
\maketitle

\begin{abstract}
Self-interacting dark matter (SIDM) models have received great attention over the past decade as solutions to the small-scale puzzles of astrophysics. Though there are different implementations of dark matter (DM) self-interactions in N-body codes of structure formation, there has not been a systematic study to compare the predictions of these different implementations. We investigate the implementation of dark matter self-interactions in two simulation codes: \Gizmo \ and \Arepo. We begin with identical initial conditions for an isolated $10^{10}$ M$_\odot$ dark matter halo and investigate the evolution of the density and velocity dispersion profiles in \Gizmo \ and \Arepo \ for SIDM cross-section over mass of 1, 5, and 50~$\rm cm^2 g^{-1}$. Our tests are restricted to the core expansion phase where the core density decreases and core radius increases with time. We find better than 30\% agreement between the codes for the density profile in this phase of evolution, with the agreement improving at higher resolution. We find that varying code-specific SIDM parameters changes the central halo density by less than 10\% outside of the convergence radius. We argue that SIDM core formation is robust across the two different schemes and conclude that these codes can reliably differentiate between cross-sections of 1, 5, and 50~$\rm cm^2 g^{-1}$ but finer distinctions would require further investigation.
\end{abstract}


%%%%%%%%%%%%%%%%%%%%%%%%%%%%%%%%%%%%%%%%%%%%%%%%%%

\section{Introduction} \label{sec:intro}

Self-interacting dark matter (SIDM) is a generic prediction of dark sector models for physics beyond the Standard Model \citep{ss2000, ahn2005, ackerman2009, arkani2009, feng2009, loeb2011, tulin2013}, and it is a possible explanation for small-scale structure formation puzzles \citep[for a comprehensive review of small-scale challenges, see][]{bullock2017}. Several groups have confirmed that SIDM with a cross-section over mass of order $1~\rm cm^2 g^{-1}$ or larger can alleviate small-scale issues \citep{Dave+2001, Colin+2002, Vogelsberger+2012, Rocha+2013, kamada2017}. This is due to the fact that scattering can effectively transfer kinetic energy within galactic halos and change the dark matter distribution \citep[see][for a review of SIDM phenomenology]{Tulin+Yu2018}. In particular, the changes introduced by the heat transfer have been shown to provide an economical way to explain the diverse range of rotation curves of galaxies \citep{oman2015, Ren+2019} and the ``too-big-to-fail" problem \citep{Vogelsberger+2012, kaplinghat2019, Turner+2021}.

Simulations of SIDM with baryons have found that while baryonic feedback can reduce the central density of a cuspy halo, if an SIDM halo already has a core present, the feedback does not make a significant difference \citep{Robles+2017, Fitts+2019, sameie2021}. This result suggests that SIDM predictions are fairly robust to feedback implementation and provides further motivation to use the observed properties of galaxies to test SIDM models.

The growing use of simulation-derived predictions to constrain the microphysics of SIDM motivates us to compare predictions from different codes. Since it is impossible to model DM-DM particle scattering directly in a galaxy formation simulation, the macroscopic effects must be modeled in an approximate way. As discussed by \citet{Tulin+Yu2018}, there are various methods for implementing DM self-interactions but the differences that may arise from each have yet to be studied in detail. Here, for the first time, we present a code comparison of two implementations of simple elastic SIDM, specifically focusing on the popular \Gizmo \ and \Arepo \ codes. We begin with identical initial conditions for an isolated $10^{10}$ M$_\odot$ dwarf halo, which is a mass regime of particular interest for small-scale structure tests. We investigate the effects of the SIDM implementations within and between the codes by varying the SIDM cross-section per mass \mbox{$\sigma$/m = 1, 5, and 50 cm$^2$ g$^{-1}$}.

Our work is structured as follows: section \ref{sec:sims} presents our initial conditions in more detail and outlines the SIDM implementations in each code. Section \ref{sec:results} presents the effects of changing the SIDM parameters, the concentration of the halo, and the resolution of the simulations. Section \ref{sec:discussion} outlines our conclusions.

%%%%%%%%%%%%%%%%%%%%%%%%%%%%%%%%%%%%%%%%%%%%%%%%%%

\section{The Simulations} \label{sec:sims}

\subsection{Code Descriptions} \label{sec:codes}

In the following, we introduce our two simulation codes -- \Gizmo \ and \Arepo \ -- and describe the methods each uses for implementing DM self-interactions.
\begin{table*}
\begin{adjustbox}{width=1\textwidth}
\begin{tabular}{lcccccccccc}

Halo name           & Mass [M$_\odot$]  & DM particles         & Run time & ($\alpha, \beta, \gamma$) & Scale radius & Force softening   & Smoothing length    & Neighbours searched    &  $\sigma$/m  & Colour \\
&&$N_p$ & [Gyr]  & & $r_s$ [kpc]& $\epsilon$ [pc] & $h_{si}$ (\Gizmo) &  $k$ (\Arepo) & [\sig] &  \\
\hline \hline
m10HR & $10^{10}$         & $5 \times 10^{6}$     & 10 & (1,3,1.536) & 14            & 10  & --       & --   & CDM          &  \crule{.25cm}{.25cm}  \\ 
 (High Resolution)  &        &       &       &     &  & & 0.25 $\epsilon$          &  32 $\pm$ 5           & 1         & \crule[darkblue]{.25cm}{.25cm}  \\
 & &   & & & & & 0.25 $\epsilon$          &  32 $\pm$ 5           & 5              & \crule[lightblue]{.25cm}{.25cm} \\
 & & & & & && 0.25 $\epsilon$          &  32 $\pm$ 5           & 50              & \crule[lightgreen]{.25cm}{.25cm}  \\

\hline

m10            & $10^{10}$         & $1 \times 10^{6}$     & 10 &  (1,3,1.536) & 14            & 10  & --       & --   & CDM          & \crule{.25cm}{.25cm}  \\

 (Baseline; &        &       &  &     &      & & 0.25 $\epsilon$          &  32 $\pm$ 5           & 1         &\crule[darkblue]{.25cm}{.25cm}   \\
 Fiducial Resolution)& &    & & & & & 0.25 $\epsilon$          &  32 $\pm$ 5           &  5              &\crule[lightblue]{.25cm}{.25cm}  \\
 & & & &  & & & 0.25 $\epsilon$          &  32 $\pm$ 5           & 50              & \crule[lightgreen]{.25cm}{.25cm}  \\
\hline
\hline
m10SIDM-      & $10^{10}$   & $1 \times 10^{6}$     & 10&  (1,3,1.536)   & 14    & 10    & 0.125 $\epsilon$ & 16 $\pm$ 5 & 5 &  \crule[darkblue]{.25cm}{.25cm}   \\
 m10SIDM+ & &    & &       & & &        
    0.5$\epsilon$ & 64 $\pm$ 5 &  5 & \crule[lightgreen]{.25cm}{.25cm} \\ 
\end{tabular}
\end{adjustbox}
\caption{Global parameters of the halos. All simulations described above have been carried out in \Gizmo \ and \Arepo. }
\label{Tab:simulationsComplete}
\end{table*}

\subsubsection{\Gizmo}    
\Gizmo \ is a massively parallel, multi-physics simulation code that uses a meshless Lagrangian Godunov-type method (``Meshless finite-mass'' or MFM; \citealt{Hopkins2015,hopkins2018}).\footnote{\Gizmo, including the SIDM module, is publicly available here: \href{https://bitbucket.org/phopkins/gizmo-public}{https://bitbucket.org/phopkins/gizmo-public}.} Given that our simulations are DM only, we rely on only the N-body tree-gravity solver which is derived from \texttt{Gadget-3} \citep{Springel2010}. 

\Gizmo's implementation of elastic self-interactions uses the methodology introduced by \citet{Rocha+2013}, which is based on the rate of scattering of the DM macro-particles in phase space. The probability of an interaction is calculated as: 
\begin{equation}
     P_{ij} = (\sigma/m) m_i v_{ij} g_{ij} \delta t,
\end{equation}
where $m_i$ is the mass of the macroparticle, $v_{ij}$ is the relative velocity difference between the two macroparticles, and $g_{ij}$ is the number density factor that accounts for the overlap of the two macroparticles' smoothing kernels. A random number is drawn from a uniform distribution in the interval $(0,1)$ to determine whether an interaction takes place. If an interaction takes place, the particles are given outgoing velocities consistent with elastic scattering. The outgoing velocities are calculated in terms of the center-of-mass velocity of the two particles, their masses, and their relative speed. The direction of the scatter is randomly chosen (such that the scatter is isotropic in the center-of-mass frame) and each particle moves opposite the other \citep[for details and tests of this implementation against analytic problems see][]{Rocha+2013}. 

\subsubsection{\Arepo}  
\Arepo \ is a massively parallel, multi-physics simulation code that employs a finite-volume method on a moving Voronoi mesh and a tree-particle-mesh method for gravitational interactions. Details of the underlying method can be found in \citep{Springel2010} while the most recent release of the code is described in \citep{Weinberger+2020}.\footnote{\Arepo \ has recently been publicly released (see \citealt{Weinberger+2020}) and is available here: \href{https://gitlab.mpcdf.mpg.de/vrs/arepo}{https://gitlab.mpcdf.mpg.de/vrs/arepo}. Note, however, that the public release does not include the SIDM implementation.}

\Arepo \ estimates the probability of an elastic self-interaction at each time step by calculating the scattering probability for each particle $i$ with each of its $k$ nearest neighbours (32 $\pm$ 5 by default and in our baseline model) $j$ as: 
\begin{equation}
    P_{ij} = (\sigma/m) m_i v_{ij} W(r_{ij},h_i)\delta t, 
\end{equation}
where $W(r_{ij}, h_i)$ is the cubic spline Kernel function and $h_i$ is the smoothing length enclosing the $k$ nearest neighbours of particle $i$ \citep[for details of this SIDM implementation, see][]{Vogelsberger+2012}.

Once the probability of interaction is calculated, a random number is drawn from a uniform distribution in the interval $(0,1)$ to determine if an interaction takes place. If it does, a neighbor $j$ must be selected to scatter with. The set of neighbors is sorted by distance to the original particle $i$ and the first neighbor $l$ whose pairwise probability to scatter with the original particle satisfies the inequality $x\leq\sum_i^l P_{ij}$ is chosen. Once a pair is matched, each particle is given a new velocity that reflects the original center-of-mass velocity and the two particles' relative velocity. The direction is chosen randomly but each particle moves opposite the other. 

Given the different approaches taken by each code in its underlying gravity solver, the similarities between their SIDM calculations and implementations are striking. With these similarities in mind, we ask: what differences, if any, arise between the two codes? 

\subsection{Initial Conditions} \label{sec:ICs}

\begin{figure}
	\includegraphics[scale=0.425]{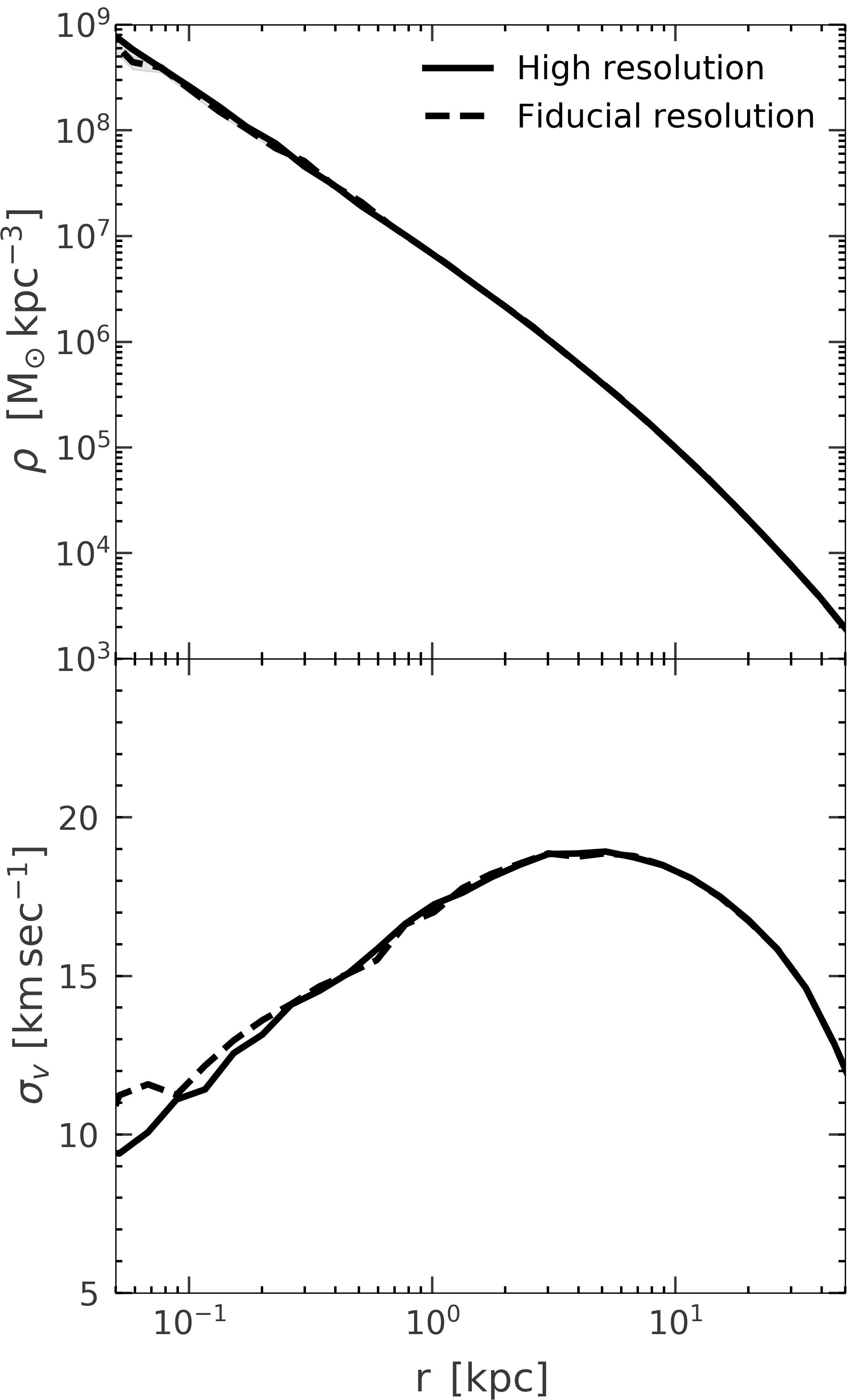}
	\centering
	\caption[sample profiles]{The density profile and velocity dispersion profile generated from \SpherIC \ for the high and fiducial resolution simulations ($5 \times 10^6$ and $1 \times 10^6$ particles respectively).}
	\label{fig:initial conditions}
\end{figure}

We use \SpherIC, an initial conditions generator for spherically symmetric systems in equilibrium first presented in \citet{SGK13}. \SpherIC \ is based on \texttt{HALOGEN4MUSE} \citep{Zemp+2008}\footnote{SpherIC was available on Bitbucket, but since that site stopped supporting Mercurial repositories, it is no longer publicly available. However \texttt{HALOGEN4MUSE} is available here: \href{https://github.com/mzemp/halogen}{https://github.com/mzemp/halogen}. Our initial conditions are also available upon request.} and generates profiles in the $(\alpha, \beta, \gamma)$-model family, a generalization of the Navarro-Frenk-White (NFW) model:
\begin{equation}
    \rho(r) = \frac{\rho_s}{(\frac{r}{r_s})^\gamma [1+(\frac{r}{r_s})^\alpha]^{(\beta-\gamma)/\alpha}} \, ,
\end{equation}
where $r_s$ is the scale radius and $\rho_s$ is the scale density \citep{navarro1997,NFW}. The parameters $\gamma$ and $\beta$ characterize the inner and outer power-law slope of the halo, respectively. The quantity $\alpha$ determines the sharpness of transition between the inner and outer slope.
An NFW profile corresponds to $(\alpha, \beta, \gamma) = (1,3,1)$ and the scale radius in this case is equal to the radius where the logarithmic slope of the density profile is -2 (i.e., $r_s = r_{-2}$). For more general ($\alpha, \beta, \gamma$)-models, we have the following relation \citep{DiCintio+2014}: 

\begin{equation}
    r_{-2} = \left( \frac{2-\gamma}{\beta-2}\right)^{1/\alpha}r_s \, .
\end{equation}

%\cite{Essig2019} have shown that high-concentration SIDM halos exhibit gravothermal core-collapse while low-concentration subhalos develop low-density cores. We therefore chose to model a more concentrated profile than an NFW to see whether \Gizmo \ and \Arepo \ would exhibit such a tendency to core-collapse and whether there would be significant differences in their results. 

In this work we utilize a halo model with $(\alpha, \beta, \gamma) = (1,3,1.536)$, $\rho_s = 1.998 \times 10^5 \mathrm{ \ M_\odot kpc^{-3}}$, and $r_s = 14.14$ kpc as these parameter values lie within the ranges that we would expect for a typical halo of this mass \citep[see][]{Lazar+2020}. We set our halos to have a total mass of $10^{10} M_\odot$. The virial radius (defined as the radius within which the average density is 100$\rho_{\textrm{crit}}$) is 52.5 kpc and the mass enclosed is $7.61\times10^9 M_\odot$. The values we use for $(\alpha, \beta, \gamma)$ correspond to $r_{-2}= 6.56$ kpc - where $r_{-2}$ is the radius at which the profile has a slope of $-2$.

We use the parameters described above to create initial conditions at two different resolution levels: a baseline (fiducial) resolution with $10^6$ particles and a higher resolution of $5\times10^6$ particles. We show the radial density and velocity dispersion profiles of these two initial conditions in Fig. \ref{fig:initial conditions}.

\subsection{Runs} \label{sec:runs}

We evolve the initial conditions at two resolutions in \Gizmo \ and \Arepo. For each resolution, we run one CDM simulation and three SIDM simulations. The results for the CDM models are shown in black in the figures throughout this work. For the code-to-code SIDM implementation comparison, we evolve both initial conditions using the respective SIDM implementations of \Gizmo \ and \Arepo \ for 3 different cross-sections: $\sigma$/m $= 1, 5, 50$ cm$^2$ g$^{-1}$. For the \Gizmo \ simulations, we adopt the default SIDM smoothing length ($h_{si}$) of 25\% of the force softening. Since the force softening we use is 10 pc, our smoothing length is 2.5 pc. Likewise, for the \Arepo \ simulations, we adopt the default value for the neighbors searched ($k$) which is 32 $\pm$ 5. 

Finally, we test the effects of varying code-specific SIDM parameters at the fiducial resolution. We set the SIDM cross-section to 5 cm$^2$ g$^{-1}$ and vary the smoothing length (in \Gizmo) and the number of neighbours searched (in \Arepo). For \Gizmo \ we adopt smoothing lengths of 1.25 pc and 5 pc which we refer to as SIDM- and SIDM+ respectively. For \Arepo \ we set the number of neighbours searched to 16 $\pm$ 5 and 64 $\pm$ 5 and refer to these again as SIDM- and SIDM+. Note that these values are less than and greater than the default values set for these parameters in the respective codes. In total, we present a suite of 20 simulations (see Table \ref{Tab:simulationsComplete} for a detailed list of all the simulations).

%%%%%%%%%%%%%%%%%%%%%%%%%%%%%%%%%%%%%%%%%%%%%%%%%%
\section{Results} \label{sec:results} 

\begin{figure*}
    \centering
    \includegraphics[scale=0.143]{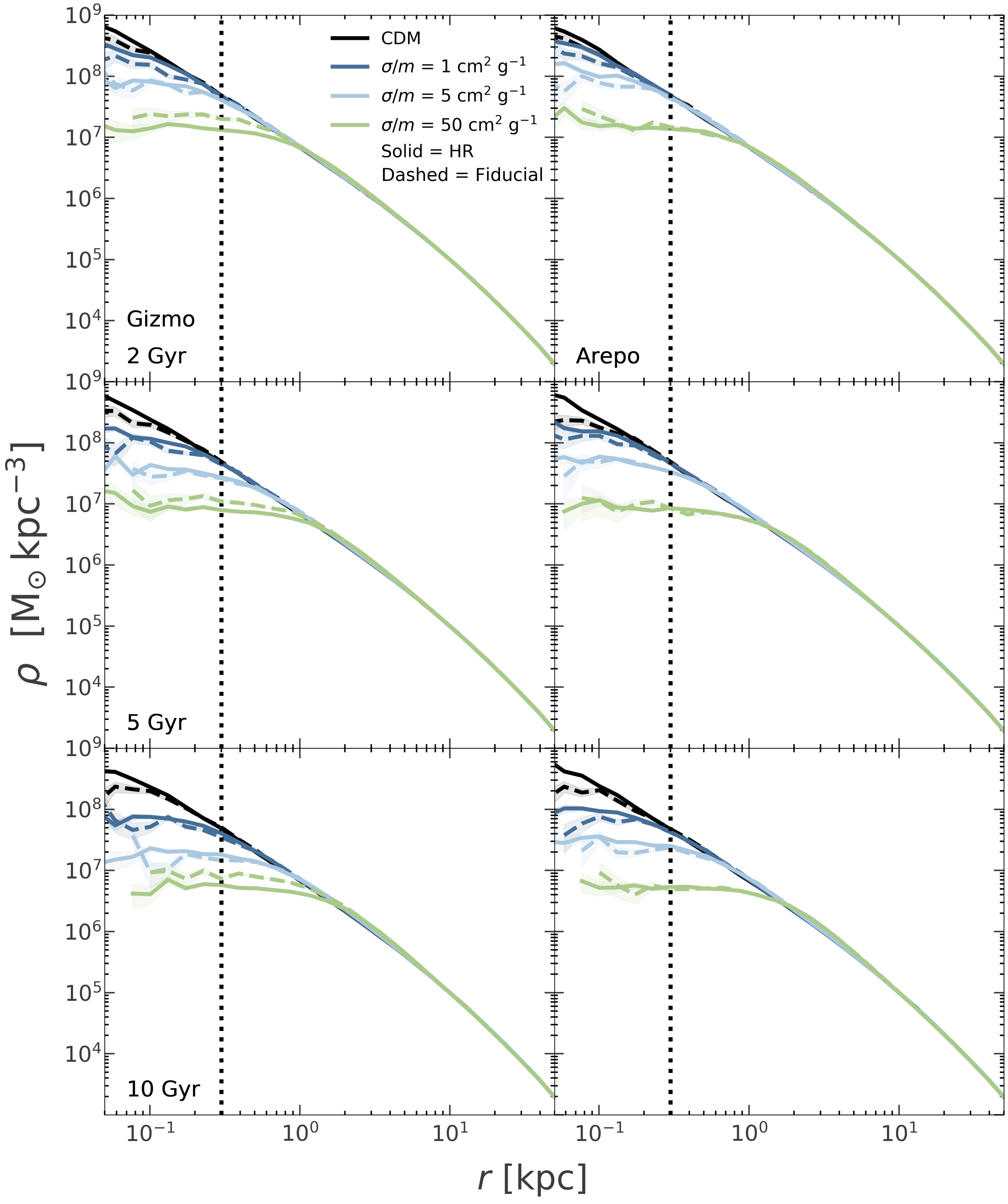}
    \caption{A resolution comparison of the two codes. The left panel depicts m10 high resolution and m10 fiducial resolution in \Gizmo \ evolved to 2, 5, and 10 Gyr. The right panel depicts the same for \Arepo. The vertical dotted line represents the radius at which the majority of our simulations are converged ($300$ pc).}
    \label{fig:intracode resolution comparison}
\end{figure*}

\subsection{Convergence of each code}
\label{sec:resolutiontest}

Before comparing the results from the two codes, we first discuss the sensitivity of each code internally to resolution. In Fig. \ref{fig:intracode resolution comparison}, we present 3 snapshots (2, 5, and 10 Gyr) of each code (\Gizmo \ in the left panel and \Arepo \ in the right). The figure shows the results of the high and fiducial resolution simulations for CDM in black and SIDM $\sigma$/m $= 1, 5$ and $50$ cm$^2$ g$^{-1}$ in dark blue, light blue, and green respectively. 

All of our high resolution simulations contain at least 200 particles within a radius of 250 pc throughout the 10 Gyr simulations. Nearly all the fiducial resolution simulations contain more than 200 particles within 300 pc.\footnote{The only exception is the fiducial resolution \Arepo \ simulation with $\sigma$/m $= 50$ cm$^2$g$^{-1}$ at late times which contains 200 particles at 400 pc.} For consistency, we mark this radius on all our plots with a black dotted line and conduct all our comparisons at 300 pc (e.g., the comparison done in Fig. \ref{fig:ratio resolution comparison}).

The error plotted in Fig. \ref{fig:intracode resolution comparison} and on all plots in this paper is the Poisson error (calculated as the density at the bin divided by the square root of the number of particles in the bin) and is represented with the shaded region around the curve. Some of the profiles are cut off because the simulations do not have any particles in the innermost region of the halo.

In sum, the simulations plotted in Fig. \ref{fig:intracode resolution comparison} demonstrate remarkable agreement between the fiducial and high resolution simulations. Indeed, outside of 250 pc, nearly all the simulations are within one another's error margins (the only exception being the $50$ cm$^2$ g$^{-1}$ run in \Gizmo). 

\subsection{Comparison of the codes across resolutions}

\begin{figure}
    \centering
    \includegraphics[scale=0.42]{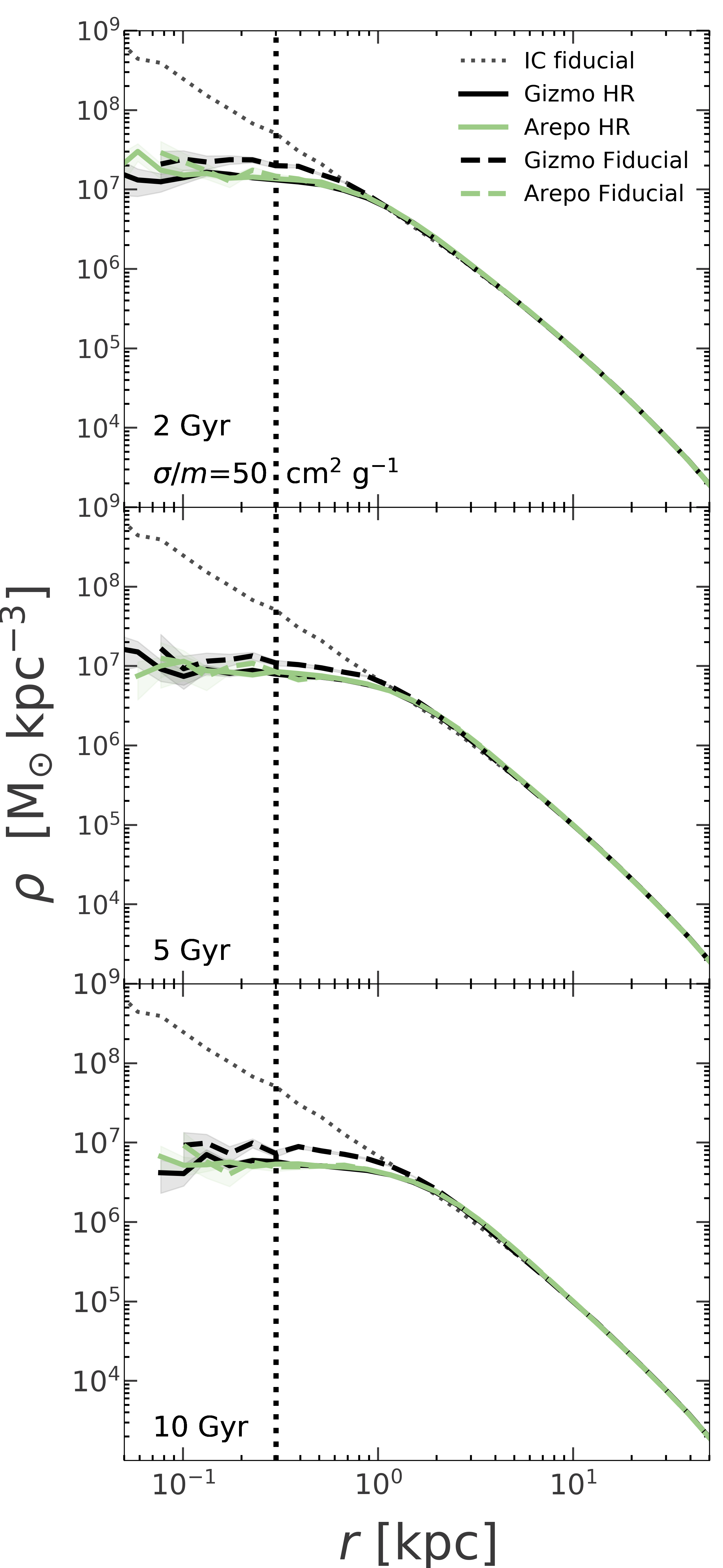} 
    \caption{From top to bottom, these figures show the halo density profiles at 2, 5, and 10 Gyr for a high resolution halo with $5 \times 10^6$ particles and $\sigma$/m = 50 cm$^2$g$^{-1}$.}
    \label{fig:m10HRPanelComparison}
\end{figure}

Having tested the convergence of the codes internally as the resolution is increased, we next compared the two resolutions across the two codes. For this comparison, we choose the most extreme SIDM cross-section ($\sigma$/m = 50 cm$^2$g$^{-1}$). Fig. \ref{fig:m10HRPanelComparison} shows the results of \Gizmo \ and \Arepo \ for this cross-section at 2, 5, and 10 Gyr. In this figure, we plot the fiducial resolution simulations in green and the high resolution in black. The \Gizmo \ simulations are plotted with solid lines while the \Arepo \ simulations are plotted with dashed lines. We find that increasing the resolution of the simulations brings the simulation results into better agreement. In other words, the density profiles from \Gizmo \ and \Arepo \ exhibit better agreement at the higher resolution than at the fiducial resolution. 

\subsection{Sensitivity to SIDM Cross-section} 

\begin{figure*}
    \centering
    \includegraphics[scale=.11]{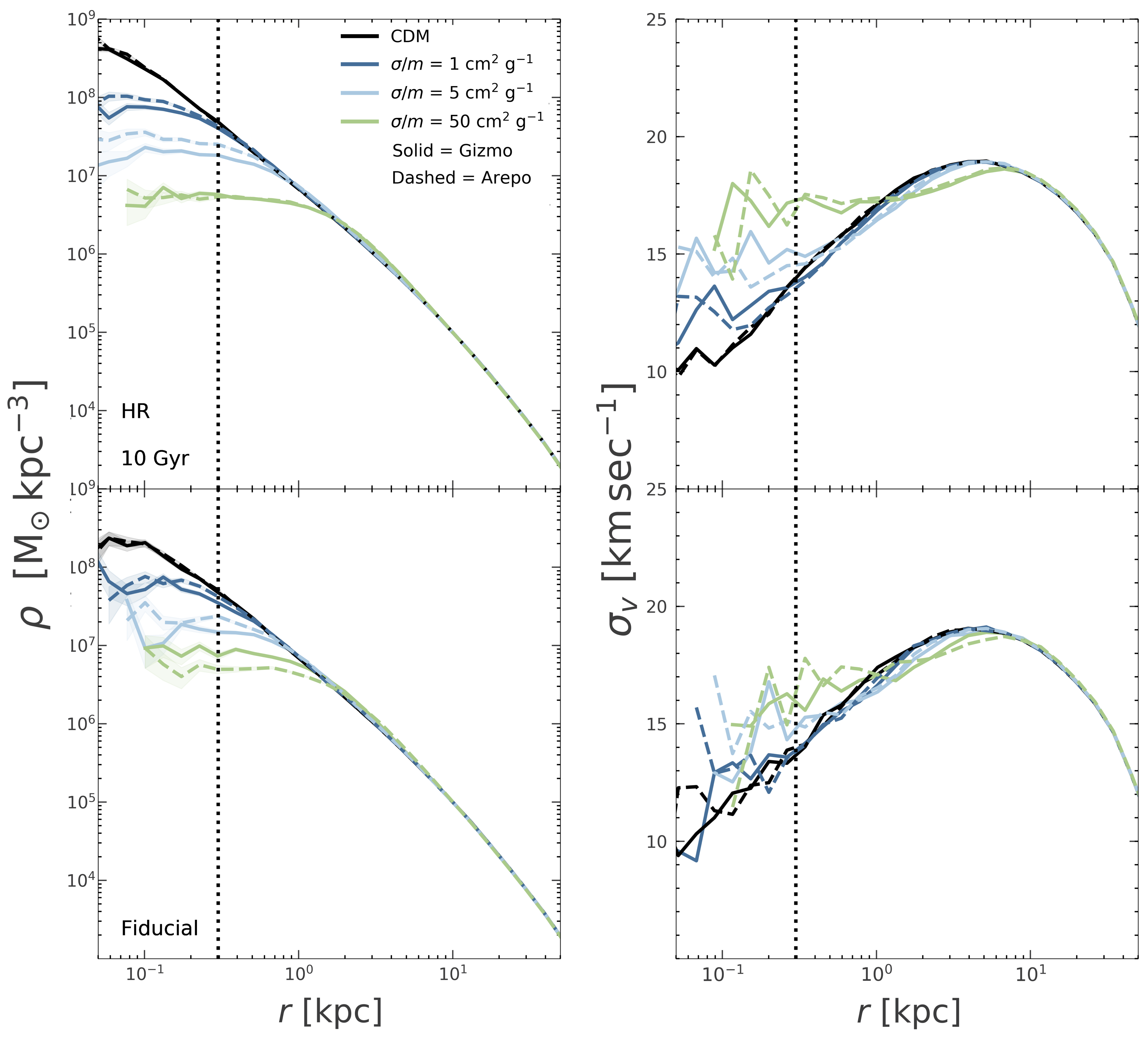}
    \caption{The top panel shows a comparison of the high resolution simulations of \Gizmo \ (solid) and \Arepo \ (dashed) evolved to 10 Gyr at various SIDM cross-sections. The bottom panel shows the same for the fiducial resolution simulation}
    \label{fig:intercode resolution comparison}
\end{figure*}

Next, we consider the sensitivity of each code to different SIDM cross-sections at high and fiducial resolution. We compare our results for the SIDM cross-section $\sigma / m$ = 1, 5, and 50 \sig \ in Fig. \ref{fig:intercode resolution comparison}. We have plotted the \Gizmo \ simulations with solid lines and the \Arepo \ simulations with dashed lines and use the same colour conventions as in Fig. \ref{fig:intracode resolution comparison}. 

Consistent with \citet{Rocha+2013}, \citet{Elbert+2015}, and  \citet{Fitts+2019}, we found that an increase in the SIDM cross-section results in density profiles that are shallower and more cored. One can better understand why the density profiles become more cored by looking at the velocity dispersion profiles. As \cite{Rocha+2013} describe, the core is created by the heat transport (characterized by the DM velocity dispersion) from large radii to the inner region. The velocity dispersion curves flattened as the halo was evolved in our simulations (see the right panel of Figure \ref{fig:intercode resolution comparison}). The flat velocity dispersion profiles of the SIDM simulations indicate that the SIDM halos are thermalized within the core, which is a necessary condition for establishing a cored DM density profile. 

\begin{figure*}
    \centering
    \includegraphics[scale=.27]{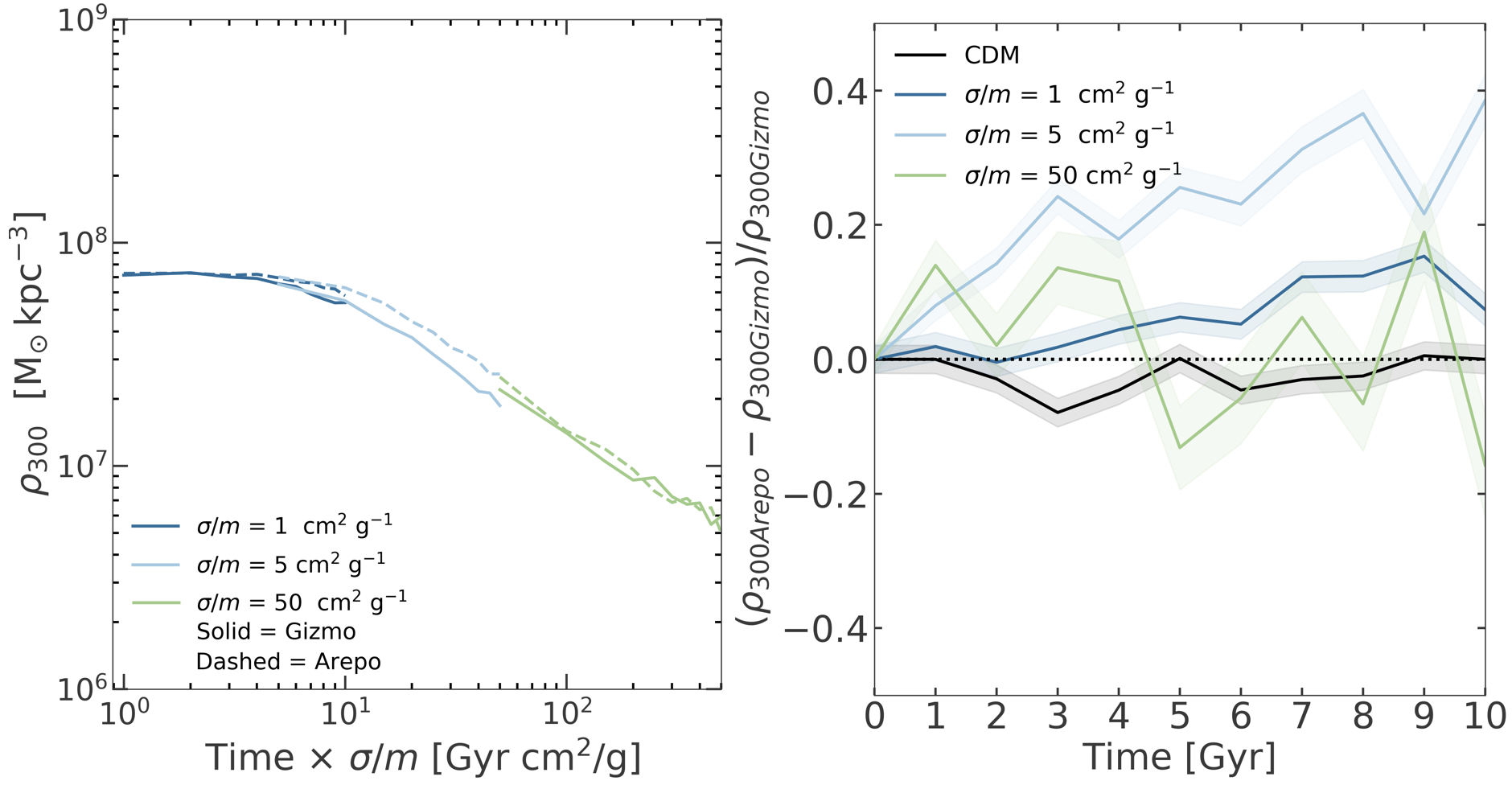}
    \caption{For the left panel, we plot the core density (at 300 pc) of each cross-section simulated as a function of the time of the simulation multiplied by the cross-section itself. For the right panel, we show the difference between the \Gizmo \ and \Arepo \ high resolution density profiles at $r=300$ pc again tracked through 10 Gyr.}
    \label{fig:ratio resolution comparison}
\end{figure*}

Analytic models of gravothermal evolution predict that halo evolution in the long-mean-free path is self-similar and determined by the product of scattering time and age \citep{KodaShapiro, Nishikawa+2020}. Given previous work (see, e.g., \cite{Ren+19} and \cite{Robertson+21}), we expect the central density for a given halo to be a function of age times $\sigma/m$ since the scattering rate is proportional to $\sigma/m$. We can see this behavior clearly in the left panel of Fig. \ref{fig:ratio resolution comparison} where we have plotted $\rho_{300}$ (proxy for the core density) vs. age multiplied by $\sigma/m$. Both \Arepo \ and \Gizmo \ follow the general trend of decreasing core density (increasing core size) with increasing age times $\sigma/m$. Note that the initial lack of evolution in $\rho_{300}$ is because the core size is smaller than $300~\rm pc$ and the density at $300~\rm pc$ is close to its initial value. 

Fig. \ref{fig:ratio resolution comparison} provides a generally encouraging picture of agreement between \Arepo \ and \Gizmo. It is worth noting that the \Arepo \ results seem to provide a more seamless curve when the three different cross-sections are plotted together as in the left panel of Fig. \ref{fig:ratio resolution comparison}, as expected from analytic models. 

To better compare the results of the two codes, we also tracked the difference between the density profiles at $r = 300$ pc through 10 Gyr. We chose $r=300$ pc since our resolution tests suggest the simulations are well-converged at this radius (see \S \ref{sec:resolutiontest} for more). For the fiducial resolution simulations, the difference in density profiles was approximately 20-40\% and the velocity dispersion profiles were within 5-10\% of one another. For the high resolution simulations, the halo profiles were largely within 30\% of one another (see the right panel of Fig. \ref{fig:ratio resolution comparison}) and the velocity dispersion profiles were within 5\%. 

We find that the difference between the two codes is much smaller than the difference between the various SIDM cross-sections we tested. In other words, we are confident these simulations can be used to distinguish amongst SIDM cross-sections of 1, 5, and 50 cm$^2$ g$^{-1}$. However, the code-to-code variation is large enough that we would argue against using such simulations to differentiate between effects due to SIDM cross-sections of, e.g., 1 vs. 1.5 cm$^2$ g$^{-1}$. 

Finally, as seen in both Fig. \ref{fig:intercode resolution comparison} and the right panel of Fig. \ref{fig:ratio resolution comparison}, there is an inversion of the DM density profiles, computed by the \Arepo \ and \Gizmo \ codes, for the SIDM cross-section $\sigma$/m $= 50 $ cm$^2 $g$^{-1}$. For cross-sections of $\sigma$/m $= 1, 5 $ cm$^2 $g$^{-1}$, the halos simulated with \Arepo \ have higher densities than those simulated with \Gizmo. However, around 4-5 Gyr, the halos with cross-sections of $50 $ cm$^2 $g$^{-1}$ evolved with \Gizmo \ become denser than the those evolved with \Arepo. The inversion is more obvious in the fiducial resolution simulation (bottom panel of Fig. \ref{fig:intercode resolution comparison}) but is also seen in the higher resolution simulations (top panel of Fig. \ref{fig:intercode resolution comparison} and right panel of Fig. \ref{fig:ratio resolution comparison}). 

To better understand the results of each code, we also tracked the number of self-interactions amongst the dark matter particles. As expected, the number of self-interactions increases with increasing cross-section. The scaling is nearly linear in both codes, with the interaction cross-section of 50 cm$^2 $g$^{-1}$ exhibiting 8 times as many interactions as the $5$ cm$^2 $g$^{-1}$ simulation and the $5$ cm$^2 $g$^{-1}$ exhibiting about 4 times as many self-interactions as the $1 $ cm$^2 $g$^{-1}$ simulation. We additionally find that the differences in the number of DM self-interactions per time step are set early on in the simulation and stay consistent throughout a 10 Gyr run. For example, the \Gizmo \ $1 $ cm$^2 $g$^{-1}$ run consistently exhibits about 10,000 DM self-interactions per Gyr throughout the entire 10 Gyr simulation and the \Arepo \ \mbox{$1 $ cm$^2 $g$^{-1}$} run consistently exhibits about 7,000 DM self-interactions per Gyr. Likewise, the \Gizmo \ $5$ cm$^2 $g$^{-1}$ run consistently has around 40,000 DM self-interactions per Gyr, and the corresponding \Arepo \ run has around 30,000.

When comparing the self-interactions between the two codes, we find that for $\sigma$/m $= 1, 5 $ cm$^2 $g$^{-1}$, \Gizmo \ exhibits a greater number of DM self-interactions ($\sim$25\% more at each time step and overall), which results in the density profiles being more cored (i.e., less dense in the inner regions). For $\sigma$/m $= 50 $ cm$^2 $g$^{-1}$, \Gizmo \ begins with slightly more self-interactions than \Arepo \ ($\sim$5\% more). However, consistent with the above discussion, there is an inversion in this trend between 3 and 5 Gyr where the number of self-interactions in \Arepo \ overtakes \Gizmo. By 6 Gyr, \Arepo \ exhibits $\sim8\%$ more self-interactions than \Gizmo \ which is maintained across the 10 Gyr.

\subsection{Sensitivity to code-specific SIDM parameters}
\begin{figure*}
    \centering
    \includegraphics[scale=0.65]{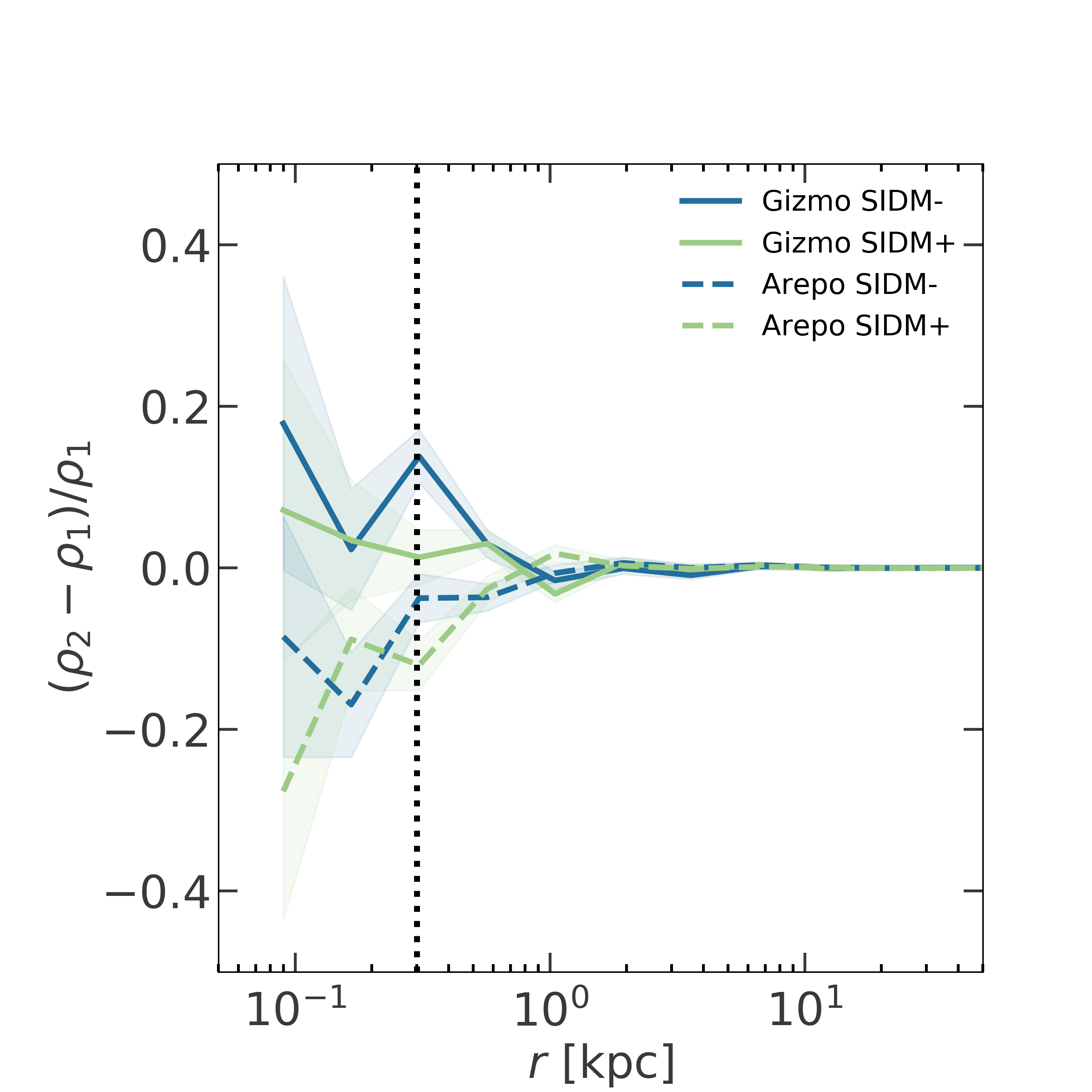}
    \caption{Sensitivity of the halo density to code-specific SIDM parameters at 5 Gyr and  $r=300$ pc for a fiducial resolution halo and $\sigma$/m = 5 cm$^2$g$^{-1}$. We vary the smoothing factor in \Gizmo \  with SIDM- corresponding to a smoothing factor of 0.125 pc and SIDM+ corresponding to a smoothing factor of 0.5 pc. Correspondingly, in \Arepo, we vary the number of neighbours searched. The SIDM- simulation corresponds to 16$\pm$5 and the SIDM+ simulation corresponds to 64$\pm$5. We then plot the difference between each of these simulations and the fiducial simulation for each code (e.g., the dark blue solid line corresponds to (\Gizmo \ SIDM-) - (\Gizmo \ fiducial))/(\Gizmo \ fiducial).}
    \label{fig:SIDMCodeSpecificComparison}
\end{figure*}

We also tested the sensitivity of each code to various code-specific SIDM parameters. In Figure \ref{fig:SIDMCodeSpecificComparison}, we plot the difference between the baseline halo density at 5 Gyr and the result of varying the smoothing factor and the number of neighbours in \Gizmo \ and \Arepo \ respectively. For \Gizmo, a smoothing factor of 25\% of the force softening is the default and has been used in, e.g., \cite{Elbert+2015}. For \Arepo, the default number of neighbours searched is 32 $\pm$ 5 but, e.g. \cite{Despali+2019} search 64 of the nearest neighbours. We therefore check the sensitivity of both codes to increasing and decreasing the default values for these parameters by a factor of 2. As seen in that figure, these code-specific SIDM choices make up to 10\% difference to the halo density at our innermost converged radius ($\sim$300 pc). However, there is no general trend with respect to the halo density apparent when the values are increased or decreased.

\section{Discussion and Conclusion} \label{sec:discussion}

Using an N-body simulation suite of an isolated dwarf dark matter halo, we compare the SIDM implementations of two simulation codes: \Gizmo \ and \Arepo. We use \SpherIC \ to generate an isolated halo with halo mass of $10^{10}\, {\rm M}_\odot$ as our initial conditions. We adopt constant elastic SIDM cross-sections $\sigma$/m $= 1, 5, {\rm \ and \ } 50$\sig \ throughout our analysis. Our main conclusions are summarized in the following. 

We find that the core density of the halo tracks the number of self-interactions predicted by the code. We find that \Gizmo \ predicts a greater number of self-interactions among the DM particles overall and that the density profiles are generally less dense in the inner region than \Arepo. These general results are consistent with previous results that have found that DM self-interactions make the halo density profile more cored (see, e.g., \citealt{Rocha+2013, Elbert+2015}). 

Increasing the resolution of the runs (from the baseline of $1\times10^6$ to $5\times10^6$ DM particles) brings the codes into better agreement. At this increased resolution, the two codes predict densities that are similar, to within 10-30\%, with Arepo consistently slightly denser than \Gizmo. Code-specific SIDM parameters made less than a 10\% difference to the final density. Generally, it seems that one can reliably predict differences between SIDM cross-sections of $\sigma$/m = 1, 5, and 50 cm$^2$g$^{-1}$ based on the agreement between these two codes. However, the inferred 30\% difference in density is large enough to preclude any finer-grained distinctions (e.g., between results obtained with with $\sigma$/m = 1.0 vs. 1.5 cm$^2$g$^{-1}$). Measuring the densities of halos may pose a larger issue than the uncertainties identified in this work.

We find that our limited scope, i.e. only comparing the SIDM implementations in each code, enables us to conduct a fruitful code comparison. In particular, instead of requiring a large comparison infrastructure and network of collaborators, we were able to isolate and discuss the effects of SIDM in the simulations. Naturally, this comparison can be extended to other codes implementing SIDM as well as to include baryonic physics and/or inelastic interactions. Having this underlying understanding of the SIDM module will be invaluable in turning to these larger projects. 

%%%%%%%%%%%%%%%%%%%% Acknowledgements %%%%%%%%%%%%%%%%%%
\section*{Acknowledgements}
The authors all thank the anonymous referee for their careful review of this manuscript. This article is partially based on work done while HM and FJM were graduate student researchers under the John Templeton Foundation grant “New Directions in Philosophy of Cosmology” (grant no. 61048). JSB and JOW also acknowledge partial support from the same grant. The opinions expressed here are those of the authors and not necessarily those of the John Templeton Foundation. FJM and JSB were also supported by National Science Foundation (NSF) grant AST-1910965. VHR acknowledges support by the Yale Center for Astronomy and Astrophysics Prize postdoctoral fellowship.

\section*{Data Availability}
The data supporting the plots within this article are available on reasonable request to the corresponding author. A public version of the \Gizmo \ code is available at \href{http://www.tapir.caltech.edu/~phopkins/Site/GIZMO.html}{http://www.tapir.caltech.edu/~phopkins/Site/GIZMO.html}. A public version of the \Arepo \ code is available at \href{https://gitlab.mpcdf.mpg.de/vrs/arepo}{https://gitlab.mpcdf.mpg.de/vrs/arepo} but we obtained a copy with the SIDM module by directly contacting the code developers.

%%%%%%%%%%%%%%%%%%%% REFERENCES %%%%%%%%%%%%%%%%%%
\bibliographystyle{mnras}
\bibliography{main.bbl} % if your bibtex file is called example.bib
 
%%%%%%%%%%%%%%%%%%%%%%%%%%%%%%%%%%%%%%%%%%%%%%%%%%

% Don't change these lines
\bsp	% typesetting comment
\label{lastpage}
\end{document}